\documentclass[journal]{IEEEtran}
\usepackage{cite}

\ifCLASSINFOpdf
  \usepackage[pdftex]{graphicx}
  \graphicspath{{../pdf/}{../jpeg/}}
  \DeclareGraphicsExtensions{.pdf,.jpeg,.png}
\else
  \usepackage[dvips]{graphicx}
  \graphicspath{{../eps/}}
  \DeclareGraphicsExtensions{.eps}
\fi

\usepackage{bm}
\usepackage{amsmath,amsfonts,stmaryrd,amssymb}

\newtheorem{remark}{Remark}
\newtheorem{theorem}{Theorem}
\newtheorem{proof}{Proof}
\newtheorem{assumption}{Assumption}

\begin{document}
%
\title{An Approach for GCI Fusion With Labeled Multitarget Densities}
%
%
%

\author{Yongwen~Jin, Jianxun~Li
\thanks{This work was jointly supported by National Natural Science Foundation (61673265); National key research and development program(2020YFC1512203);Special research projects for civil aircraft(MJ-2017-S-38); Shanghai Commercial Aircraft System Engineering Joint Research Fund ; CEMEE (2019K0302A).}}

%
%

\markboth{Journal of \LaTeX\ Class Files,~Vol.~14, No.~8, August~2015}%
{Shell \MakeLowercase{\textit{et al.}}: Bare Demo of IEEEtran.cls for IEEE Journals}
%



\maketitle

\begin{abstract}
This paper addresses the Generalized Covariance Intersection (GCI) fusion method for labeled random finite sets. We propose a joint label space for the support of fused labeled random finite sets to represent the label association between different agents, avoiding the label consistency condition for the label-wise GCI fusion algorithm. Specifically, we devise the joint label space by the direct product of all label spaces for each agent. Then we apply the GCI fusion method to obtain the joint labeled multi-target density. The joint labeled RFS is then marginalized into a general labeled RFS, providing that each target is represented by a single Bernoulli component with a unique label. The joint labeled GCI (JL-GCI) for fusing LMB RFSs from different agents is demonstrated. We also propose the simplified JL-GCI method given the assumption that targets are well-separated in the scenario. The simulation result presents the effectiveness of label inconsistency and excellent performance in challenging tracking scenarios.
\end{abstract}


%
\IEEEpeerreviewmaketitle

\section{Introduction}

The multi-source multi-target tracking problem has increasingly received attention in the signal processing community. When the information from multiple sources is transferred to an agent or a fusion center, in order to obtain a more precise estimation compared with that of the single sensor, it is important to design a reasonable strategy to combine the information. Theoretically, if the information is represented by the measurement sets, the computational complexity for fusion will increase exponentially with the number of sensors \cite{Fantacci2016}. Usually, the measurements from each sensor are processed sequentially in a heuristic method \cite{Mahler2010, Liu2016}. If the information is the multi-target densities, Chong et al. \cite{Chong1990} presented the optimal solution to this problem, but the computational costs can be too expensive for real-time estimation. Some sub-optimal distributed multi-source fusion algorithm, like the Cauchy-Schwarz divergence based fusion method or the weighted arithmetic average (WAA) method \cite{Gostar2017}-\cite{Yi2020}, the Generalized Covariance Intersection (GCI) or the Kullback-Leibler average (KLA) \cite{Battistelli2015}-\cite{Battistelli2013}, have been proposed for this issue, where the GCI method is discussed in this paper. GCI method is immune to the double-counting effect for multi-sensor fusion and usually has higher estimation accuracy than the WAA method, but leas to consistency fusion of multiple densities \cite{Yi2020, Wang2018}. Based on GCI method, the fusion algorithms for PHD/CPHD \cite{Battistelli2015,Battistelli2013,Lai2019,Li2017}, multi-Bernoulli filter \cite{Wang2017}-\cite{Yi2020_2} are proposed. However, these algorithms do not involve the fusion of labeled RFS, thus generating no fused tracks directly.

Some GCI fusion methods for labeled RFS have been proposed for track-to-track fusion. \cite{Fantacci2018} proposed an analytical GCI fusion method for labeled multi-Bernoulli densities and M$\delta$-GLMB densities. Assume that all agents share the same label space, the analytical solution of the GCI fusion can be obtained. However, the labeled multi-sensor GCI fusion is sensitive to the label inconsistencies among different agents. \cite{Li2018} analyzed this issue and proposed the label inconsistency indicator and the "yes-object probability" to quantify the label inconsistencies among multi-target posteriors from different agents. Based on the analysis, the labeled multi-target posteriors are first marginalized to the unlabeled posteriors, then the GCI fusion is applied. Li et al. \cite{Li2019} proposed the LM-GCI method, in which the cost matrix is first constructed based on the label inconsistency indicator, then the Hungarian algorithm \cite{Heskes1998, Kuhn2010} is used for finding the optimal label matching $\tau^*$, the label-wise GCI fusion can be performed with $\tau^*$. If the number of nodes in the sensor network is $S$, the largest cardinality of the label space among the agents is $L_{max}$ and the Bernoulli component for the LMB RFS is characterized by $M$ Gaussian components, the overall computational complexity for LM-GCI is $\mathcal{O}(\max\{SL_{max}^2M^2,S L_{max}\})$.

We propose an approach for the GCI method induced by the principle of RFS and FISST directly, solving the problem of label inconsistency. The main contribution of this paper is the following four parts:

First, we devise the joint label space by the direct product of multiple label spaces. Every element in the joint label space represents the labels for all agents and the association among labels without additional representations. The joint label space is an extension of the standard label space with a major difference: more than one label may point to the same target. This difference leads to different distinct label indicator in the joint label space, which will be further discussed in this paper.

Second, we derive the joint labeled GCI fusion for the LMB RFS based on minimizing the weighted Kullback-Leibler divergence. Specifically, it can be seen from the weighted KL divergence, the fused multi-target density function is proportional to the weighted multiply of the multi-target densities generated by each sensor agent. By normalization, we can obtain the multi-target density, which is indeed the M$\delta$-GLMB RFS. The difficulty of the method is to obtain the normalization factor. In this paper, we use Murty's algorithm \cite{Murty1968, Miller1997} to obtain several state sets with the largest unnormalized coefficients as the approximation of the multi-target state space, then the normalization factor is the summation over these unnormalized coefficients. In this case, the joint labeled GCI (JL-GCI) method is obtained with a higher computational complexity but better estimation accuracy compared with LM-GCI.

Third, we provide an approximate method under the condition of well-separated targets. If the targets are separable, in other words, the distance between any two targets in the field of view is large, the unnormalized coefficients corresponding to those unsatisfying the label uniqueness approaches 0. By introducing these unqualified but negligible components, the analytical solution for GCI fusion with labeled multi-Bernoulli (LMB) random finite set (RFS) can be obtained. The simplified JL-GCI method has lower computational complexity than that of LM-GCI and is capable of dealing with label inconsistency.

Forth, we demonstrate the proposed algorithm, named the JL-GCI method, is equivalent to the LM-GCI method given the optimal label matching. The label matching $\tau$ can be denoted as a subset of the joint label space. The LM-GCI is regarded as a method of implementing GCI fusion under an optimal subset in the joint label space. If the joint label space is restricted to a subset of it, two methods are identical.

The rest of the paper is organized as follows. The multi-sensor multi-target distributed fusion method is briefly introduced in Section \ref{Background}. The label inconsistency problem within the labeled multi-sensor multi-target fusion and corresponding solutions are demonstrated in Section \ref{LabelInconsistencyProblemAndExistingSolutions}. The definition of joint label space, two GCI fusion methods, and the relationship between proposed methods and LM-GCI are proposed in Section \ref{GCIFusionWithJointLabelSpace}. The implementations for the two algorithms are presented in Section \ref{ImplementationForJLGCIAndSimplifiedJLGCI}. The simulation result is presented in Section \ref{Simulation}. The conclusion is provided in Section \ref{Conclusion}.

\section{Background}
\label{Background}

In this section, the RFS based distributed fusion for multi-target densities is introduced, including the problem of distributed fusion \cite{Mahler2010, Battistelli2015, Mahler2000, Fantacci2018, Khaleghi2013}, RFS theory\cite{Reuter2014}-\cite{Ristic2016}, GCI fusion for labeled and unlabeled multi-target densities. 

\subsection{Problem Description}

For the multi-target tracking problem, suppose that there are $S$ agents in the sensor network, each agent receives measurements denoted by a random finite set \cite{Mahler2003},
\begin{equation}
    Z_{i}^k=\{z_{1,i}^k,\ldots,z_{m_i,i}^k\},
\end{equation}
where $m_i$ is the number of measurements received by $i$th sensor, denoted by the cardinality of a finite set. It is worth noting that not all elements of the measurement set come from the targets in the field of view (FoV), usually including some clutters with a known distribution. The number of clutter is usually assumed to be Poisson, while the spatial distribution is uniform \cite{Reuter2014}-\cite{Vo2009}.

Given the distribution of newborn targets (a multi-target density function or intensity function) and measurement set, the multi-target posterior can be obtained by each sensor locally, which is represented by the multi-target density function $f_i(X)$ or the first-order statistical moment (probability hypothesis density) $D_i(x)$. $X$ is a random finite set, representing the multi-target state:
\begin{equation}
    X_{i}^k=\{x_1,\ldots,x_{n_k}\}.
\end{equation}
Specifically, $X$ may be a multi-Bernoulli RFS, LMB RFS, or $\delta$-GLMB RFS.

In the sensor network, each agent locally updates multi-target information utilizing the multi-target dynamics and the local measurement set, transfers the information to a specific center or exchanges the information with other agents, then applies the fusion method to combine available information. More formally, the distributed multi-target tracking problem over the network can be stated as follows. Each node $i=1,\ldots,S$ uses tracking algorithm to integrate into a multi-target density of $X$ based on all local measurements from $1$ to $k$. The multi-target density of $X$ is then combined with those from other agents to form an integral multi-target density $f(X)$, rendering the fusion method for $f(X)$ is guaranteed to be optimal in a certain sense \cite{Battistelli2013}.

The main issue for distributed multi-target tracking problem is the way of devising a fusion algorithm, i.e., how to integrate information from different agents, rather than devising the single-sensor multi-target tracking algorithm. The design of the fusion method can be investigated with respect to the categories of the multi-target densities, like the probability hypothesis density, the multi-Bernoulli density, the labeled multi-Bernoulli density, or the generalized labeled multi-Bernoulli density. Some requirements for the method, like the scalability for which the computational costs of each agent must be independent of the size of the network \cite{Battistelli2013}, or limited FoV for each sensor, are considered. In this paper, we focus on devising a general fusion method for labeled multi-target densities.

\subsection{Labeled Multi-target Density}

A labeled state is a single target state $x\in\mathbb{X}$ augmented by the label $\ell\in\mathbb{L}$, i.e. $\bm{x}=(x,\ell)$, where $\ell$ calibrates the identity of the state for extracting track in multi-target tracking scenario. The label is usually extracted from a discrete label set $\mathbb{L}=\{ \alpha_i:i\in\mathbb{N} \}$, where all elements are different. For each target, the label can be represented by $\ell=(k,i)$, where $k$ represents the birth-time of a target, $i\in\mathbb{N}$ is a unique ordinal number to distinguish targets the simultaneous newborn targets \cite{Reuter2014}.

Throughout the paper, we follow the same notation adopted in \cite{Vo2013, Li2019}: Single-target states and corresponding measurements are represented by lowercase letters, e.g., $x$, $\bm{x}$, $z$. While the multi-target states and measurement sets are represented by uppercase letters, e.g. $X$, $\bm{X}$, $Z$. The multi-target densities are represented by the function of sets, e.g. $f(X)$, $\pi(X)$, $\bm{f}(\bm{X})$, $\bm{\pi}(\bm{X})$. Symbols for labeled states and their distributions are bolded to distinguish them from unlabeled ones, e.g. $\bm{x}$, $\bm{X}$, $\bm{f}$, $\bm{\pi}$. Spaces are represented by blackboard bold, e.g., $\mathbb{X}$, $\mathbb{Z}$, $\mathbb{N}$, $\mathbb{L}$, while $\mathcal{F}(\mathbb{X})$ represents the collection of finite subsets of $\mathbb{X}$ and $\mathbb{F}_n(\mathbb{X})$ represents the collection of finite subsets of $\mathbb{X}$ with $n$ elements.

The label set of a labeled RFS $\bm{X}$ is represented by $\mathcal{L}(\bm{X})=\{ \mathcal{L}(\bm{x}):\bm{x}\in\bm{X} \}$, where $\mathcal{L}:\mathbb{X}\times\mathbb{L}\to\mathbb{L}$ is a mapping defined by $\mathcal{L}((x,\ell))=\ell$. Define the distinct label indicator:
\begin{equation}
    \Delta(\bm{X})=\delta_{|\bm{X}|}(|\mathcal{L}(\bm{X})|).
\end{equation}

The multi-target density is represented by the GLMB RFS:
\begin{equation}
\label{eq:GLMB}
    \bm{f}(\bm{X})=\Delta(\bm{X})\sum_{c\in\mathbb{C}}w^{(c)}(\mathcal{L}(\bm{X}))[p^{(c)}]^{\bm{X}},
\end{equation}
where $\mathbb{C}$ is a discrete index set, $w^{(c)}(L)$ and $p^{(c)}$ satisfy
\begin{equation}
    \sum_{L\subseteq\mathbb{L}}\sum_{c\in\mathbb{C}}w^{(c)}(L)=1,
\end{equation}
\begin{equation}
    \int p^{(c)}(x,\ell)dx = 1.
\end{equation}
For any labeled RFS, the corresponding unlabeled version can be obtained by marginalizing the labeled multi-target density:
\begin{equation}
\label{eq:LabelMarginalized1}
    f(\{x_1,\ldots,x_n\})=\sum_{(\ell_1,\ldots,\ell_n)\in\mathbb{L}^n}\bm{f}(\{(x_1,\ell_1),\ldots,(x_n,\ell_n)\}).
\end{equation}

If $\mathbb{C}$ is singleton, (\ref{eq:GLMB}) can be simplified:
\begin{equation}
\label{eq:LMB1}
    \bm{f}(\bm{X})=\Delta(\bm{X})w(\mathcal{L}(\bm{X}))p^{\bm{X}}.
\end{equation}
If $p^{\bm{X}}$ is rewritten by $[p(\cdot;\mathcal{L}(\bm{X}))]^{\bm{X}}$, (\ref{eq:LMB1}) is the M$\delta$-GLMB RFS \cite{Fantacci2016, Fantacci2018}. If
\begin{equation}
    w(L)=\prod_{\ell\in\mathbb{L}}(1-r(\ell))\prod_{\ell'\in L}\frac{1_{\mathbb{L}(\ell')r(\ell')}}{1-r(\ell')},
\end{equation}
(\ref{eq:LMB1}) is the LMB RFS, which can be represented by $\{r(\ell),p(x,\ell)\}_{\ell\in\mathbb{L}}$.

\subsection{GCI Fusion}

The GCI fusion, generalized by the weighted fusion for two Gaussian densities, is proposed for the distributed multi-sensor multi-target tracking problem \cite{Mahler2000}. Suppose the agent $i=1,\ldots,S$ generates the multi-target density $f_i(X)$ based on the local measurements and multi-target dynamics, then the GCI fusion is the geometric mean of the local multi-target densities,
\begin{equation}
    f_{\omega}(X)=\frac{\prod_{i=1}^S[f_i(X)]^{\omega_i}}{\int \prod_{i=1}^S[f_i(X)]^{\omega_i} \delta X},
\end{equation}
where $\int\cdot\delta X$ is the set integral, $\omega_i$ is the averaging weight satisfying $\omega_i\geq 0,i=1,\ldots,S$ and $\sum_{i=1}^S\omega_i=1$.

The GCI fusion is also denoted by the Kullback-Leibler Average (KLA) fusion since the method can be obtained by minimizing weighted KL divergence:
\begin{equation}
    f_{KLA}(X)\triangleq \arg\min_{f}\sum_{i}\omega_i D_{KL}(f\|f_i),
\end{equation}
\begin{equation}
    D_{KL}(f\|f_i)\triangleq \int f(X)\log{\frac{f(X)}{f_i(X)}}\delta X,
\end{equation}
$f_{KLA}(X)$ is identical to the formula of the GCI fusion \cite{Battistelli2013,Heskes1998}. Thus, we will the term GCI and KLA interchangeably in this paper.

\subsection{Label-wise GCI Fusion}
\label{LabelwiseGCIFusion}
Suppose the labeled multi-target states belong to $\mathcal{F}(\mathbb{X}\times\mathbb{L})$ for all agents in the sensor network, Fantacci et al. \cite{Fantacci2018} presented that the GCI fusion for LMB RFS can be obtained analytically. Specifically, the KLA of the LMB densities $\bm{f}_i=\{ (r_i(\ell), p_i(\cdot,\ell)) \}_{\ell\in\mathbb{L}}$ given the normalized nonnegative weights $\omega_i,i=1,\ldots,S$, is the LMB RFS:
\begin{equation}
    \bar{\bm{f}}=\{(\bar{r}(\ell),\bar{p}(\cdot,\ell))\}_{\ell\in\mathbb{L}},
\end{equation}
where
\begin{align}
    \bar{r}(\ell)&=\frac{\int\prod_{i=1}^S(r_i(\ell)p_i(x,\ell))^{\omega_i}dx}{\prod_{i=1}^S(1-r_i(\ell))^{\omega_i}+\int\prod_{i=1}^S(r_i(\ell)p_i(x,\ell))^{\omega_i}dx},\\
    \bar{p}(\cdot,\ell)&=\frac{\prod_{i=1}^S p_i(\cdot,\ell)^{\omega_i}}{\int\prod_{i=1}^S p_i(x,\ell)^{\omega_i}dx}.\label{eq:GCI4SingleTargetDensity}
\end{align}

The analytical solution of GCI fusion is of informative contribution, but the assumption of label consistency usually does not hold. The labels with respect to different agents are probably diverse, especially when the types of sensors are different. Even if the way for generating labels is the same, it still cannot guarantee label consistency since the labels for every agent are generated locally and separately. Therefore, the way of dealing with the label inconsistency is vital for the labeled multi-sensor multi-target fusion method.

\section{Label Inconsistency Problem And Existing Solutions}
\label{LabelInconsistencyProblemAndExistingSolutions}

The GCI fusion method proposed in section \ref{LabelwiseGCIFusion} needs the assumption that all LMB densities from different agents share the same label space. However, this assumption is not available, since the definition of the label for different agents can be diverse, or the existence of clutters and miss-detection affects the generation of labels.

In this section, the label inconsistency indicator \cite{Li2019,Li2018} and the metric for applying label-wise GCI fusion by minimizing the indicator are introduced.

\subsection{Label Inconsistency Indicator}

Given an unlabeled multi-target state $x_1,\ldots,x_n$, the labelling information is conditional joint probability distribution of their corresponding labels $\ell_1,\ldots,\ell_n$,
\begin{equation}
    \varpi(\{ (\ell_1|x_1),\ldots,(\ell_n|x_n) \})=\frac{\bm{\pi}(\{ (x_1,\ell_1),\ldots,(x_n,\ell_n) \})}{\pi(\{ x_1,\ldots,x_n \})},
\end{equation}
where $\pi(\{x_1,\ldots,x_n\})$ is the unlabeled version of $\bm{\pi}(\cdot)$, which can be obtained according to (\ref{eq:LabelMarginalized1}).

Then, define the inconsistency of labelling information for a given set of unlabeled states $x_1,\ldots,x_n$ according to the multiple conditional multi-label distributions $\varpi_i(\{(\ell_1|x_1),\ldots,(\ell_n|x_n)\}),i=1,\ldots,S$ and the GCI coefficients, 
\begin{equation}
    \mu_{\bm{\Pi}}(\{x_1,\ldots,x_n\})=\sum_{(\ell_1,\ldots,\ell_n)\in\mathbb{L}^n}\prod_{i=1}^S[\varpi_i(\{ (\ell_1|x_1),\ldots,(\ell_n|x_n)\})]^{\omega_i}.
\end{equation}

The label inconsistency indicator with respect to different multi-target densities is
\begin{equation}
    d_G(\bm{\Pi})\triangleq -\log{\mathbb{E}_{\pi_\omega}[\mu_{\bm{\Pi}}(X)]}.
\end{equation}

\subsection{Label Matching GCI Fusion}

The LF-GCI \cite{Li2018} is proposed for dealing with the label inconsistency problem, in which the labeled multi-target posteriors are first marginalized to their unlabeled versions then applying the GCI fusion. Li et al. \cite{Li2019} proposed the LM-GCI method. In this method, the optimal problem according to the label inconsistency indicator is proposed for obtaining the optimal label matching $\tau^*$. Then perform the label-wise GCI fusion of $\bm{\pi}_a(\bm{X})$ and $\bm{\pi}_b(\bm{X})$. Here we mainly introduce the LM-GCI method.

Consider two nodes $a$, $b$ in the sensor network with the label spaces $\mathbb{L}_a$ and $\mathbb{L}_b$ respectively. Define the bijective label matching $\tau:\mathbb{L}_a\to\mathbb{L}_b$. The collection for all label matching is $\mathcal{T}(\mathbb{L}_a,\mathbb{L}_b)$. For any subset $I\subseteq\mathbb{L}_a$, define $\tau(I)\triangleq \cup_{\ell\in I}\tau(\ell)$.

Given the label matching $\tau$, the labeled multi-target state of node $b$ can be expressed as a labeled multi-target state of node $a$,
\begin{equation}
\begin{aligned}
    &\bm{\pi}_b^{(\tau)}(\{ (x_1,\ell_1),\ldots,(x_n,\ell_n) \})\\
    &=\bm{\pi}_b(\{ (x_1,\tau(\ell_1)),\ldots,(x_n,\tau(\ell_n)) \}).
\end{aligned}
\end{equation}
Therefore, if the optimal label matching $\tau^*$ is obtained, the label inconsistency problem can be solved by mapping different label spaces to one specific space. Here $\tau^*$ can be obtained by minimizing $d_G(\{ (\bm{\pi}_a,\omega_a),(\bm{\pi}_b^{(\tau)},\omega_b) \})$,
\begin{equation}
\label{eq:LabelMatchingOptimal1}
    \tau^*=\arg\min_{\tau\in\mathcal{T}(\mathbb{L}_a,\mathbb{L}_b)}d_G(\{ (\bm{\pi}_a,\omega_a),(\bm{\pi}_b^{(\tau)},\omega_b) \}).
\end{equation}

Specifically, by utilizing the Hungarian algorithm \cite{Kuhn2010} with a cost matrix, the optimal matching $\tau^*$ is estimated. Assumed that $\bm{\pi}_a$ and $\bm{\pi}_b$ are LMB RFSs characterized by the following parameter sets,
\begin{equation}
    \bm{\pi}_a=\{ (r_a(\ell),p_a(\cdot,\ell)) \}_{\ell\in\Psi_a},
\end{equation}
\begin{equation}
    \bm{\pi}_b = \{ (r_b(\ell),p_b(\cdot,\ell)) \}_{\ell'\in\Psi_b}.
\end{equation}
Without loss of generality, the number of elements in $\Psi_a$ is supposed larger than that of $\Psi_b$, that is, $|\Psi_a|>|\Psi_b|$. Consider an auxiliary label set $\Psi_b^A$ constructed as
\begin{equation}
    \Psi_b^A=\Psi_b\cup\Psi_b^C,
\end{equation}
where $\Psi_b^C$ is an arbitrary subset of $\mathbb{L}_b \backslash\Psi_b$. The cost function is constructed by
\begin{equation}
    \bm{C}=\begin{bmatrix}
    C_{\ell_1,\ell_1'} &\cdots & C_{\ell_1,\ell_{|\Psi_a|}'}\\
    \vdots &\ddots &\vdots\\
    C_{\ell_{|\Psi_a|},\ell_1'} &\cdots &C_{\ell_{|\Psi_a|},\ell_{\Psi_a}'}
    \end{bmatrix},
\end{equation}
where
\begin{equation}
    C_{\ell,\ell'}=\left\{
    \begin{aligned}
    &\begin{aligned}-\log[(1-r_a(\ell))^{\omega_a} (1-r_b(\ell'))^{\omega_b}\\
    + (r_a(\ell))^{\omega_a}(r_b(\ell'))^{\omega_b}c_{\ell,\ell'}]\end{aligned},&&\ell\in\Psi_b\\
    &-\log{(1-r_a(\ell))^{\omega_a}},&&\ell'\in\Psi_b^C,
    \end{aligned}
    \right.
\end{equation}
with
\begin{equation}
    c_{\ell,\ell'}=\int p_a(x,\ell)^{\omega_a}p_b(x,\ell')^{\omega_b}dx.
\end{equation}

The association matrix is obtained by the Hungarian algorithm \cite{Kuhn2010}
\begin{equation}
    \bm{S}\triangleq \begin{bmatrix}
    S_{\ell_1,\ell_1'} &\cdots &S_{\ell_1,\ell_{|\Psi_a|}'}\\
    \vdots &\ddots &\vdots\\
    S_{\ell_{|\Psi_a|},\ell_1'} &\cdots &S_{\ell_{|\Psi_a|},\ell_{|\Psi_a|}'}
    \end{bmatrix}
\end{equation}
where
\begin{equation}
    S_{\ell,\ell'}\triangleq\left\{
    \begin{aligned}
    &1,&& \text{if }\ell'=\tau(\ell),\\
    &0,&&\text{otherwise.}
    \end{aligned}
    \right.
\end{equation}
Then, the fused multi-target density can be obtained by label-wise GCI method.

For the label matching with respect to multiple agents (usually the number of agents is lager than 2), the solution among agents is to seek the label matchings in a pair-wise way, i.e., to perform (\ref{eq:LabelMatchingOptimal1}) pair-wise. Given that there are $S$ agents in the sensor network, the LMB RFS for the $i$th agent includes $\| \Psi_i \|$ Bernoulli components. Furthermore, if we apply the Gaussin mixture model with $M$ components, the computational complexity is $\mathcal{O}(\max{\{ SL_{max}^2M^2,SL_{max}^3 \}})$, where $L_{max}=\max_{i=1}^S|\Psi_i|$.

The LM-GCI method introduces the matching mechanism to solve the label inconsistency problem. The simulation result in \cite{Li2019} provides the efficiency of the algorithm. In this paper, we attempt to deal with this issue on the basis of the RFS theory, i.e., to incorporate the label inconsistency to the RFS based fusion method. The LM-GCI can be considered as the GCI method combining with an optimal subset searching method.

\section{GCI Fusion With Joint Label Space}
\label{GCIFusionWithJointLabelSpace}

In this section, we propose the GCI method with the joint label space. The joint label space is constructed by the direct product for all label spaces of the agents. Hence, every element in the joint label space represents a possible label matching between different spaces. Then we obtain the joint labeled multi-target density based on label-wise GCI fusion and change it into the LMB RFS. Last, the multi-target state is extracted from the LMB RFS.

First, we propose the definition of the joint label space and the difference compared with the classic label space. Then, according to the analysis of the fusion method, we obtain the joint labeled GCI (JL-GCI) method. Finally, we propose two fusion algorithms: the first one is the application for the JL-GCI fusion method with respect to the LMB RFS, while the second one is an approximation GCI method given the assumption that the targets in the scenario are well separated. In the last part of this section, we will demonstrate the relationship between the proposed algorithms and the LM-GCI method. Besides, the application of the JL-GCI method with respect to GLMB and M$\delta$-GLMB is also discussed.

\subsection{Joint Label Space}

Suppose there are $S$ agents in the sensor network, all agents deliver the LMB RFS $\{(r(\ell_{j,i}),p(x_j,\ell_{j,i}))\}_{\ell_{j,i}\in\mathbb{L}_i}$ to the fusion center (or a specific agent), here $\mathbb{L}_i$ is the label space corresponding to the $i$th agent. We construct the joint label space:
\begin{equation}
    \mathbb{L}_1\times\mathbb{L}_2\times\cdots\times\mathbb{L}_S:=\mathbb{L}.
\end{equation}
The label set of a labeled RFS $\bm{X}\subseteq \mathbb{X}\times\mathbb{L}$ is
\begin{equation}
    \mathcal{L}(\bm{X})=\{(\ell_{1,1},\ldots,\ell_{1,S}),\ldots,(\ell_{n,1},\ldots,\ell_{n,S})\},
\end{equation}
Let
\begin{equation}
\label{eq:Joint2Single}
    \bm{X}(i)=\{(x_1,\ell_{1,i}),\ldots,(x_n,\ell_{n,i})\},
\end{equation}
here $\bm{X}(i)$ is the projection in the labeled state space $\mathbb{X}\times\mathbb{L}_i$. If $L\subseteq\mathbb{L}$, then
\begin{equation}
    L(i)=\{ \ell_{1,i},\ldots,\ell_{n,i} \},
\end{equation}

For any real valued function $h_i$ defined in $\mathbb{X}\times\mathbb{L}_i$, where $i=1,\ldots,S$, let
\begin{equation}
    [ \prod_{i=1}^S h_i ]^{\bm{X}}=\prod_{i=1}^S[h_i]^{\bm{X}(i)}=\prod_{i=1}^S\prod_{(x,\ell_i)\in\bm{X}(i)}h_i(x,\ell_i).
\end{equation}

\begin{remark}
It is worth noting that the span over the domains of $h_i(i=1,\ldots,S)$ is the domain of $\bm{X}$, thus the definition is reasonable. $\bm{f}(\bm{X}(i))$ can be obtained by marginalizing $\bm{f}(\bm{X})$,
\begin{equation}
\label{eq:MarginalizedLabel}
\begin{small}
\begin{aligned}
    &\bm{f}(\{(x_1,\ell_{1,i}),\ldots,(x_n,\ell_{n,i})\})\\
    &=\sum_{(\ell_{1,1},\ldots,\ell_{n,1})\in\mathbb{L}_1^n}\cdots\sum_{(\ell_{1,i-1},\ldots,\ell_{n,i-1})\in\mathbb{L}_{i-1}^n}\sum_{(\ell_{1,i+1},\ldots,\ell_{n,i+1})\in\mathbb{L}_{i+1}^n}\\
    &\cdots\sum_{(\ell_{1,S},\ldots,\ell_{n,S})\mathbb{L}_S^{n}}\bm{f}(\{(x_1,\ell_{1,1},\ldots,\ell_{1,S}),\cdots,(x_n,\ell_{n,1},\ldots,\ell_{n,S})\})\\
    &=\bm{f}(\bm{X}(i)).
\end{aligned}
\end{small}
\end{equation}

Similarly, if $f:\mathbb{L}^n\to\mathbb{R}$ is symmetric,
\begin{equation}
\label{eq:jointLabelSpaceSum2}
\begin{aligned}
    &\sum_{(\ell_{1,1},\ldots,\ell_{n,1})\in\mathbb{L}_1^n}\cdots\sum_{(\ell_{1,S},\ldots,\ell_{n,S})\in\mathbb{L}_S^n}\\
    &\delta_n(|\{ (\ell_{1,1},\ldots,\ell_{1,S}),\ldots,(\ell_{n,1},\ldots,\ell_{n,S}) \}|)\cdot\\
    &f((\ell_{1,1},\ldots,\ell_{1,S}),\ldots,(\ell_{n,1},\ldots,\ell_{n,S}))\\
    &=n!\sum_{\{\ell_1,\ldots,\ell_n\}\in\mathcal{F}_n(\mathbb{L})}f((\ell_{1,1},\ldots,\ell_{1,S}),\ldots,(\ell_{n,1},\ldots,\ell_{n,S})).
\end{aligned}
\end{equation}
\end{remark}

\begin{remark}
(\ref{eq:jointLabelSpaceSum2}) is not equal to
\begin{equation}
    n!\sum_{L\in\mathcal{F}(\mathbb{L})}f(L).
\end{equation}
The latter only requires that $L$ be the $n$ different labels in $\mathbb(L)$
\begin{equation}
(\ell_{i,1},\ldots,\ell_{i,S})\neq(\ell_{j,1},\ldots,\ell_{j,S}),\quad i\neq j,    
\end{equation}
which is not necessarily satisfied
\begin{equation}
    \delta_n(|L(i)|)=1, \quad i = 1,\ldots,S.
\end{equation}

\end{remark}

\subsection{Joint Labeled GCI Fusion}

The multi-sensor multi-target distributed fusion method can be obtained by solving an optimization problem, like the LM-GCI method, or the GCI method proved by minimizing the KL divergence. In this section, we derive the joint labeled GCI fusion method. Specifically, we first define the joint label space by the direct sum of the label spaces of all agents in the sensor network, then the joint labeled GCI fusion method is obtained by minimizing the weighted KL divergence.

Given the labeled multi-target probability density function $\bm{f}_i(\cdot)$ of the $i$th agent with the label space $\mathbb{X}\times\mathbb{L}_i$, where $i=1,\ldots,S$, the fused multi-target probability density function is $\bm{f}(\bm{X})$, where
\begin{equation}
\begin{aligned}
    \bm{X}&=\{(x_1,\ell_{1,1},\ldots,\ell_{1,S}),\ldots,(x_n,\ell_{n,1},\ldots,\ell_{n,S})\}\\
    &\subseteq \mathbb{X}\times \mathbb{L}_1\times\cdots\times\mathbb{L}_S\triangleq\mathbb{X}\times\mathbb{L}.
\end{aligned}
\end{equation}
That is, we obtain the labeled multi-target state by combining the label space for each agent, which is the domain of the fused labeled multi-target state density. The joint label space is different from any label space of the agent, but the elements in the joint labeled state space can be projected to any labeled space of the agent based on (\ref{eq:Joint2Single}).

According to the definition of KL divergence, if the fused multi-target distribution density is $\bm{f}(\bm{X})$, the multi-target distribution density for each agent is $\bm{f}_i(\bm{X}(i))$, then the weighted KL divergence is
\begin{equation}
\begin{aligned}
    &\sum_{i=1}^S \omega_i D_{KL}(\bm{f};\bm{f}_i)=\sum_{i=1}^S\omega_i[\int \bm{f}(\bm{X})\log{\bm{f}(\bm{X})}\delta\bm{X}\\
    &- \int \sum_{j\neq i: L_j\in\mathbb{L}_j^n}^S \bm{f}(\bm{X})\log\bm{f}_i(\bm{X}(i))\delta \bm{X}(i) ]\\
    &=\int \bm{f}(\bm{X})\log{ \frac{\bm{f}(\bm{X})}{ [\prod_{i=1}^S \bm{f}_i^{\omega_i}]^{\bm{X}} } }\delta\bm{X}.
\end{aligned}
\end{equation}
Let
\begin{equation}
\label{eq:KLA0}
    \bar{\bm{f}}(\bm{X})=\frac{[\prod_{i=1}^S \bm{f}_i^{\omega_i}]^{\bm{X}}}{C},
\end{equation}
where
\begin{equation}
\label{eq:NormalizationCoefficient}
    C = \int [\prod_{i=1}^S \bm{f}_i^{\omega_i}]^{\bm{X}}\delta\bm{X}.
\end{equation}
Therefore, the weighted KL divergence is given by
\begin{equation}
\begin{aligned}
    \sum_{i=1}^S D_{KL}(\bm{f};\bm{f}_i)&=\int \bm{f}(\bm{X})\log{ \frac{\bm{f}(\bm{X})}{ C\cdot \bar{\bm{f}}(\bm{X}) } }\delta\bm{X}\\
    &=D_{KL}(\bm{f};\bar{\bm{f}})-\log{C}.
\end{aligned}
\end{equation}
Since the KL divergence is non-negative and the minimization is obtained if and only if $\bm{f}=\bar{\bm{f}}$, then the KLA is given by
\begin{equation}
\label{eq:GCIFusion4JointLabelSet}
    \bm{f}_V(\bm{X}) = \arg\min_{\bm{f}}\sum_{i=1}^S D_{KL}(\bm{f};\bm{f}_i)=\frac{[\prod_{i=1}^S \bm{f}_i^{\omega_i}]^{\bm{X}}}{ \int [\prod_{i=1}^S \bm{f}_i^{\omega_i}]^{\bm{X}}\delta\bm{X} },
\end{equation}

The focus of the fusion method is the denominator of the fraction, the normalization factor $C$. Theoretically, the set integral of the denominator needs to solve the integral over all labeled multi-target states, which is usually intractable. Therefore, in this paper, we attempt to solve this issue through two methods. One method is to directly select a number of multi-target states (finite sets) to represent $\mathcal{F}(\mathbb{X}\times\mathbb{L})$. Hence, the more multi-target states are selected, the more accurate the fused multi-target density function is. Another method is to simplify the equation by introducing an approximation, thus obtaining the analytical solution of the fusion method.

\subsection{GCI Fusion For Labeled Multi-Bernoulli Densities}

\begin{theorem}
Given the multi-target densities $\bm{f}_i(\bm{X}(i))$ for all agents in the sensor network, where $i=1,\ldots,S$,
\begin{equation}
\begin{aligned}
    \bm{f}_i(\bm{X}(i))&=\Delta(\bm{X}(i))\prod_{\ell\in\mathbb{L}_i}(1-r_i(\ell))\\
    &\prod_{(x,\ell)\in\bm{X}(i)}\frac{1_{\mathbb{L}_i}(\ell)r_i(\ell)p_i(x,\ell)}{1-r_i(\ell)}.
\end{aligned}
\end{equation}
The fused multi-target density is a GLMB RFS, 
\begin{equation}
\label{eq:fusedDistribution1}
    \bm{f}_V(\bm{X})=\prod_{i=1}^S\Delta(\bm{X}(i))w(\mathcal{L}(\bm{X}))[\bar{p}(\cdot)]^{\bm{X}},
\end{equation}
where
\begin{equation}
    w(\mathcal{L}(\bm{X}))=\frac{1}{C}\cdot [\prod_{i=1}^S(1-r_i)^{\omega_i}]^{\mathbb{L}\backslash\mathcal{L}(\bm{X})}[\eta\prod_{i=1}^S r_i^{\omega_i}]^{\mathcal{L}(\bm{X})},
\end{equation}
\begin{equation}
    \eta(\ell_{j,1},\ldots,\ell_{j,S})=\int \prod_{i=1}^S p_i(x,\ell_{j,i})^{\omega_i} dx,
\end{equation}
\begin{equation}
\begin{aligned}
    C = \int \prod_{i=1}^S\Delta(\bm{X}(i))[\prod_{i=1}^S (1-r_i)^{\omega_i}]^{\mathbb{L}\backslash\mathcal{L}(\bm{X})}[\eta\prod_{i=1}^S r_i^{\omega_i}]^{\mathcal{L}(\bm{X})}\delta\bm{X},
\end{aligned}
\end{equation}
$\bar{p}(\cdot)$ is given by (\ref{eq:GCI4SingleTargetDensity}).
\end{theorem}

\begin{proof}
According to the multi-target density generated by $i$th agent,
\begin{equation}
\label{eq:KLA1}
\begin{aligned}
    &\prod_{i=1}^S \bm{f}_i(\{(x_1,\ell_{1,i}),\cdots,(x_n,\ell_{n,i})\})^{\omega_i}\\
    &=\prod_{i=1}^S\Delta(\bm{X}(i))[\prod_{\ell\in\mathbb{L}_i\backslash\{\ell_{1,i},\ldots,\ell_{n,i}\}}(1-r_i(\ell))\\
    &\quad\prod_{j=1}^n 1_{\mathbb{L}_i}(\ell_{j,i})r_i(\ell_{j,i})p_i(x_j,\ell_{j,i})]^{\omega_i},
\end{aligned}
\end{equation}
Since $\ell_{j,i}\in\mathbb{L}_i$, $1_{\mathbb{L}_i(\ell_{j,i})}$ can be negligible, let
\begin{align*}
    \prod_{i=1}^S[\prod_{\ell\in\mathbb{L}_i\backslash\{\ell_{1,i},\ldots,\ell_{n,i}\}}(1-r_i(\ell))]^{\omega_i}&=[\prod_{i=1}^S(1-r_i)^{\omega_i}]^{\mathbb{L}\backslash\mathcal{L}(\bm{X})},\\
    \prod_{i=1}^S[\prod_{j=1}^n r_i(\ell_{j,i}) ]^{\omega_i}&=[\prod_{i=1}^S r_i^{\omega_i}]^{\mathcal{L}(\bm{X})},\\
    \prod_{i=1}^S[\prod_{j=1}^np_i(x_j,\ell_{j,i})]^{\omega_i}&=[\prod_{i=1}^S (p_i)^{\omega_i}]^{\bm{X}}
\end{align*}
Thus (\ref{eq:KLA1}) can be rewritten by
\begin{equation}
\label{eq:KLA2}
\begin{small}
\begin{aligned}
    &\prod_{i=1}^S \bm{f}_i(\{(x_1,\ell_{1,i}),\ldots,(x_n,\ell_{n,i})\})^{\omega_i}\\
    &=\prod_{i=1}^S\Delta(\bm{X}(i))[\prod_{i=1}^S (1-r_i)^{\omega_i}]^{\mathbb{L}\backslash\mathcal{L}(\bm{X})}[\prod_{i=1}^S r_i^{\omega_i}]^{\mathcal{L}(\bm{X})}[\prod_{i=1}^S (p_i)^{\omega_i}]^{\bm{X}}.
\end{aligned}
\end{small}
\end{equation}
Let
\begin{equation}
    \eta(\ell_{j,1},\ldots,\ell_{j,S})=\int \prod_{i=1}^S p_i(x,\ell_{j,i})^{\omega_i} dx,
\end{equation}
\begin{equation}
\label{eq:singleTargetFusion}
    \bar{p}(x,\ell_{j,1},\ldots,\ell_{j,S})=\frac{\prod_{i=1}^S p_i(x,\ell_{j,i})^{\omega_i} }{\eta(\ell_{j,1},\ldots,\ell_{j,S})},
\end{equation}
(\ref{eq:KLA2}) can be rewritten by
\begin{equation}
\label{eq:KLA3}
\begin{aligned}
    &\prod_{i=1}^S \bm{f}_i(\{(x_1,\ell_{1,i}),\ldots,(x_n,\ell_{n,i})\})^{\omega_i}\\
    &=\prod_{i=1}^S\Delta(\bm{X}(i))[\prod_{i=1}^S (1-r_i)^{\omega_i}]^{\mathbb{L}\backslash\mathcal{L}(\bm{X})}[\eta\prod_{i=1}^S r_i^{\omega_i}]^{\mathcal{L}(\bm{X})}[\bar{p}]^{\bm{X}}.
\end{aligned}
\end{equation}
Substitute (\ref{eq:KLA3}) into (\ref{eq:NormalizationCoefficient}), we have
\begin{equation}
\begin{aligned}
    C = \int \prod_{i=1}^S\Delta(\bm{X}(i))[\prod_{i=1}^S (1-r_i)^{\omega_i}]^{\mathbb{L}\backslash\mathcal{L}(\bm{X})}[\eta\prod_{i=1}^S r_i^{\omega_i}]^{\mathcal{L}(\bm{X})}\delta\bm{X}.
\end{aligned}
\end{equation}

Therefore, the fused multi-target density is a GLMB RFS,
\begin{equation}
    \bm{f}_V(\bm{X})=\prod_{i=1}^S\Delta(\bm{X}(i))w(\mathcal{L}(\bm{X}))[\bar{p}]^{\bm{X}},
\end{equation}
where
\begin{equation}
    w(\mathcal{L}(\bm{X}))=\frac{1}{C}\cdot [\prod_{i=1}^S(1-r_i)^{\omega_i}]^{\mathbb{L}\backslash\mathcal{L}(\bm{X})}[\eta\prod_{i=1}^S r_i^{\omega_i}]^{\mathcal{L}(\bm{X})}
\end{equation}
\end{proof}

The normalization coefficient $C$ in the fused multi-target density can be replaced by the summation over several largest wights of the multi-target states. Specifically, for the sensor network with two agents, a cost matrix can be constructed, and $n$ state sets with the largest weight can be obtained by Murty's algorithm. The construction of the cost matrix will be introduced in section \ref{ImplementationForTwoSensors}, the description of Murty's algorithm is referred to \cite{Murty1968, Miller1997}.

\subsection{Analytical Approximation}

\begin{assumption}
\label{Assumption1}
The targets in the scenario are well separated, i.e., for any two agents $a$, $b$, if $\ell_{j,a}$ and $\ell_{j,b}$ point to different targets,
\begin{equation}
    p_{a}(x,\ell_{j,a})^{\omega_{a}}p_{b}(x,\ell_{j,b})^{\omega_{b}}\approx 0,\quad\forall x\in\mathbb{X}.
\end{equation}
If there is label mismatching among the joint labeled multi-target state $\bm{X}$, then the corresponding weight is
\begin{equation}
\label{eq:assumption1}
    w(\mathcal{L}(\bm{X}))=0.
\end{equation}
\end{assumption}
\begin{remark}
If there is an assignment error in the joint labeled multi-target state $\bm{X}$, e.g., for a specific joint label $(\ell_{j,1},\ldots,\ell_{j,S})$, the label $\ell_{j,i_a}$ of the $i_a$th agent and the label $\ell_{j,i_b}$ of the $i_b$th agent correspond to different target, thus we have
\begin{equation}
    p_{i_a}(x,\ell_{j,i_a})^{\omega_{i_a}}p_{i_b}(x,\ell_{j,i_b})^{\omega_{i_b}}=0,\quad\forall x\in\mathbb{X}.
\end{equation}
According to the definition of $\eta(\cdot)$, there is
\begin{equation}
    \eta(\ell_{j,1},\ldots,\ell_{j,S})=0.
\end{equation}
Thus (\ref{eq:assumption1}) holds true.
\end{remark}

\begin{theorem}
\label{thm:Approximation1}
Given the Assumption \ref{Assumption1}, the fused multi-target density function can be approximated to
\begin{equation}
\label{eq:approximation1}
    \bm{f}_V(\bm{X})=\Delta(\bm{X})w(\mathcal{L}(\bm{X}))[\bar{p}]^{\bm{X}},
\end{equation}
where
\begin{equation}
    \Delta(\bm{X})=\delta_{n}(|\{ (\ell_{1,1},\ldots,\ell_{1,S}),\ldots,(\ell_{n,1},\ldots,\ell_{n,S}) \}|),
\end{equation}
\begin{equation}
    w(L)=\prod_{\ell\in\mathbb{L}}(1-\bar{r}(\ell))\prod_{\ell'\in L}\frac{\bar{r}(\ell')}{1-\bar{r}(\ell')},
\end{equation}
\begin{equation}
\label{eq:bar_r1}
\begin{aligned}
    &\bar{r}(\ell)=\bar{r}(\ell_1,\ldots,\ell_S)\\
    &=\frac{\eta(\ell_1,\ldots,\ell_S)\prod_{i=1}^S[r_i(\ell_i)]^{\omega_i}}{\prod_{i=1}^S [1-r_i(\ell_i)]^{\omega_i} + \eta(\ell_1,\ldots,\ell_S)\prod_{i=1}^S[r_i(\ell_i)]^{\omega_i}}.
\end{aligned}
\end{equation}
\end{theorem}

The difference between (\ref{eq:approximation1}) and (\ref{eq:fusedDistribution1}) is the definition of the distinct label indicator. For $(\ell_{i,1},\ldots,\ell_{i,S})$ and $(\ell_{j,1},\ldots,\ell_{j,S})$, (\ref{eq:approximation1}) requires 
\begin{equation}
\label{eq:DistinctLabelIndicator1}
    \exists s\in\{1,\ldots,S\},\quad\text{s.t. }\ell_{i,s}\neq\ell_{j,s},
\end{equation}
While (\ref{eq:fusedDistribution1}) requires
\begin{equation}
\label{eq:DistinctLabelIndicator2}
    \forall s\in\{1,\ldots,S\},\quad\text{s.t. }\ell_{i,s}\neq\ell_{j,s}.
\end{equation}

To prove Theorem \ref{thm:Approximation1}, we can state that, given the Assumption \ref{Assumption1}, for any cases that satisfy (\ref{eq:DistinctLabelIndicator1}) but do not satisfy (\ref{eq:DistinctLabelIndicator2}), the weight $w(\mathcal{L}(\bm{X}))$ of the multi-target state set $\bm{X}$ is 0. While the proof of the analytical expression of $w(L)$ is similar to the derivation in \cite{Fantacci2018}.

\begin{proof}
For any joint label pair $(\ell_{i,1},\ldots,\ell_{i,S})$, $(\ell_{j,1},\ldots,\ell_{j,S})$, satisfying (\ref{eq:DistinctLabelIndicator1}) but not (\ref{eq:DistinctLabelIndicator2}), suppose
\begin{equation}
\label{eq:unreasonableCase}
    \left\{
    \begin{aligned}
        \ell_{i,s_a}\neq\ell_{j,s_a},\\
        \ell_{i,s_b}=\ell_{j,s_b},
    \end{aligned}
    \right.
\end{equation}
for the $s_a$th agent and the $i_b$th agent. If the assignment between $\ell_{i,s_a}$ and $\ell_{i,s_b}$ is error, based on the Assumption \ref{Assumption1}, we have
\begin{equation}
    \eta(\ell_{i,1},\ldots,\ell_{i,S})=0,
\end{equation}
thus (\ref{eq:assumption1}) holds true.

If the assignment between $\ell_{i,s_a}$ and $\ell_{i,s_b}$ is correct, two labels are assigned to the same target. Since $\ell_{i,s_a}\neq\ell_{j,s_a}$, $\ell_{i,s_a}$ and $\ell_{j,s_a}$ are assigned to different targets; Since $\ell_{i,s_b}=\ell_{j,s_b}$, $\ell_{i,s_b}$ and $\ell_{j,s_b}$ are assigned to the same target. Thus $\ell_{j,s_a}$ and $\ell_{j,s_b}$ are assigned to different targets. Then we get
\begin{equation}
    \eta(\ell_{j,1},\ldots,\ell_{j,S})=0,
\end{equation}
thus (\ref{eq:assumption1}) holds true.

Given the Assumption \ref{Assumption1}, we have
\begin{equation}
\label{eq:KLA4}
\begin{aligned}
    &\prod_{i=1}^S \bm{f}_i(\{(x_1,\ell_{1,i}),\ldots,(x_n,\ell_{n,i})\})^{\omega_i}\\
    &=\Delta(\bm{X})[\prod_{i=1}^S (1-r_i)^{\omega_i}]^{\mathbb{L}\backslash\mathcal{L}(\bm{X})}[\prod_{i=1}^S r_i^{\omega_i}]^{\mathcal{L}(\bm{X})}[\prod_{i=1}^S (p_i)^{\omega_i}]^{\bm{X}}.
\end{aligned}
\end{equation}
The normalization coefficient $C$ is given by
\begin{equation}
\begin{aligned}
    C=\sum_{L\in\mathcal{F}(\mathbb{L})}[\prod_{i=1}^S (1-r_i)^{\omega_i} ]^{\mathbb{L}\backslash L}[\eta\prod_{i=1}^S(r_i)^{\omega_i}]^L,
\end{aligned}
\end{equation}
According to the Lemma 3 in \cite{Vo2013},
\begin{equation}
    C = [ \prod_{i=1}^S(1-r_i)^{\omega_i} + \eta\prod_{i=1}^S(r_i)^{\omega_i} ]^{\mathbb{L}}.
\end{equation}
Therefore, the fused posterior multi-target density function is given by
\begin{equation}
\begin{aligned}
    \bm{f}_V(\bm{X})&=\frac{[\prod_{i=1}^S \bm{f}_i^{\omega_i}]^{\bm{X}}}{ \int [\prod_{i=1}^S \bm{f}_i^{\omega_i}]^{\bm{X}}\delta\bm{X} }\\
    &=\Delta(\bm{X})[\frac{\prod_{i=1}^S(1-r_i)^{\omega_i}}{\prod_{i=1}^S (1-r_i)^{\omega_i} + \eta\prod_{i=1}^S(r_i)^{\omega_i} }]^{\mathbb{L}\backslash\mathcal{L}(\bm{X})}\\
    &\quad [\frac{\eta\prod_{i=1}^S(r_i)^{\omega_i}}{\prod_{i=1}^S (1-r_i)^{\omega_i} + \eta\prod_{i=1}^S(r_i)^{\omega_i}}]^{\mathcal{L}(\bm{X})}[\bar{p}]^{\bm{X}}\\
    &=\Delta(\bm{X})\prod_{\ell\in\mathbb{L}}(1-\bar{r}(\ell))\prod_{\ell'\in\mathcal{L}(\bm{X})}\frac{\bar{r}(\ell')}{1-r(\ell')}[\bar{p}]^{\bm{X}}\\
    &=\Delta(\bm{X})w(\mathcal{L}(\bm{X}))[\bar{p}]^{\bm|{X}}.
\end{aligned}
\end{equation}
\end{proof}
Though this method introduces a strong assumption, the computational cost is lower compared with JL-GCI and LM-GCI.

\subsection{Discussions}

Consider the label spaces $\mathbb{L}_1$ and $\mathbb{L}_2$ for two agents, if there is a mapping $\tau:\mathbb{L}_1\to\mathbb{L}_2$ between two label spaces, the mapping can be expressed by the set,
\begin{equation}
\label{eq:LabelAssociationSet}
    \tau=\{(\ell_1,\ell_2):\ell_1\in\mathbb{L}_1,\ell_2\in\mathbb{L}_2\}.
\end{equation}
The set (\ref{eq:LabelAssociationSet}) is a subset of the joint label set, i.e.
\begin{equation}
    \tau\subseteq\mathbb{L},\quad \text{where }\mathbb{L}=\mathbb{L}_1\times\mathbb{L}_2.
\end{equation}
If the joint label set is restrained to $\tau$, the joint labeled multi-target density (\ref{eq:fusedDistribution1}) is equal to (\ref{eq:approximation1}), since (\ref{eq:unreasonableCase}) is not likely to happen. Therefore, the JL-GCI and the simplified JL-GCI output the same fused multi-target density, which is equal to that of the LM-GCI. In this case, we consider the LM-GCI as a particular approach in the context of the joint labeled GCI fusion: first, reduce the joint label space to the label space of a particular agent then perform GCI fusion.

Though the GCI method proposed in this paper is associated with LMB RFS. However, the JL-GCI method is not restrained to a particular density. One can obtain the GCI method with respect to other labeled multi-target densities. 

Given the M$\delta$-GLMB RFS generated by $S$ agents in the sensor network,
\begin{equation}
\label{eq:MdeltaGLMB1}
\begin{aligned}
    \bm{f}_i(\bm{X}(i))&=\Delta(\bm{X}(i))w(\mathcal{L}(\bm{X}(i)))\prod_{(x,\ell)\in\bm{X}(i)}p(x,\ell;\mathcal{L}(\bm{X}(i)))\\
    &=\Delta(\bm{X}(i))w(\mathcal{L}(\bm{X}(i)))[p(\cdot;\mathcal{L}(\bm{X}(i)))]^{\bm{X}(i)}.
\end{aligned}
\end{equation}
Substitute (\ref{eq:GCIFusion4JointLabelSet}) into (\ref{eq:MdeltaGLMB1}),
\begin{equation}
\begin{aligned}
    \bm{f}_V(\bm{X})&=\frac{\prod_{i=1}^S\{\Delta(\bm{X}(i))w(\mathcal{L}(\bm{X}(i)))[p(\cdot;\mathcal{L}(\bm{X}(i)))]^{\bm{X}(i)}\}^{\omega_i}}{C}\\
    &=\prod_{i=1}^S\Delta(\bm{X}(i))\cdot w(\mathcal{L}(\bm{X}))[\bar{p}]^{\bm{X}},
\end{aligned}
\end{equation}
where
\begin{equation}
    C=\int \prod_{i=1}^S\{\Delta(\bm{X}(i))w(\mathcal{L}(\bm{X}(i)))[p(\cdot;\mathcal{L}(\bm{X}(i)))]^{\bm{X}(i)}\}^{\omega_i} \delta \bm{X},
\end{equation}
\begin{equation}
    w(L)=\frac{1}{C}\prod_{i=1}^S w(L)\int \prod_{i=1}^S[\prod_{\ell\in L(i)}p_i(x,\ell;L(i))]^{\omega_i}dx.
\end{equation}

\section{Implementation For JL-GCI And Simplified JL-GCI}
\label{ImplementationForJLGCIAndSimplifiedJLGCI}

In this section, we discuss the realization of the joint labeled GCI fusion method and its simplified version. For the case of combining two labeled multi-target densities, we first construct the cost matrix, then obtaining the computational procedure of JL-GCI and its simplified version. In the second part of this section, we analyze the computational complexity of the algorithms.

\subsection{Implementation for Two Sensors}
\label{ImplementationForTwoSensors}

The realization of the joint labeled GCI fusion method is divided into four steps:
\begin{enumerate}
    \item Construct the cost matrix $\bm{C}$.
    \item Applying Murty's algorithm to obtain the best $k$ multi-target states with minimum cost, then obtaining the corresponding weights.
    \item Obtaining the fused multi-target density function by normalization.
    \item For the convenience of target indexing, we marginalize the joint label to the standard label by (\ref{eq:MarginalizedLabel}).
\end{enumerate}

Given the LMB RFSs $\bm{f}_a(\bm{X})=\{(r_a(\ell),p_a(x,\ell))\}_{\ell\in\Psi_a}$ and $\bm{f}_b(\bm{X})=\{(r_b(\ell),p_b(x,\ell))\}_{\ell\in\Psi_b}$ generated by agents $a$ and $b$ respectively, $\Psi_a\subseteq\mathbb{L}_a$ and $\Psi_b\subseteq\mathbb{L}_b$ are the support of $\bm{f}_a$ and $\bm{f}_b$ respectively, For the convenient of demonstration, let $|\Psi_a|=\psi_a$, $|\Psi_b|=\psi_b$,
\begin{equation}
    \Psi_a=\{ \ell_{1,a},\ldots,\ell_{\psi_a,a} \},
\end{equation}
\begin{equation}
    \Psi_b=\{ \ell_{1,b},\ldots,\ell_{\psi_b,b} \}.
\end{equation}

Define the cost matrix $\bm{C}_1$ by
\begin{equation}
    \bm{C}_1=\begin{bmatrix}
    C_{1,1}^{(1)} &\cdots &C_{1,\psi_b}^{(1)}\\
    \vdots &\ddots &\vdots\\
    C_{\psi_a,1}^{(1)} &\cdots &C_{\psi_a,\psi_b}^{(1)}
    \end{bmatrix},
\end{equation}
where
\begin{equation}
    C_{i,j}^{(1)}=-\log[\frac{\eta(\ell_{i,a},\ell_{j,b}) r_a(\ell_{i,a})^{\omega_a} r_b(\ell_{j,b})^{\omega_b} }{(1-r_a(\ell_{i,a}))^{\omega_a}(1-r_b(\ell_{j,b}))^{\omega_b}}].
\end{equation}
Define the cost matrix $\bm{C}_2$ by
\begin{equation}
    \bm{C}_2=\begin{bmatrix}
    0 & &\infty\\
     &\ddots &\\
     \infty & &0
    \end{bmatrix}_{\psi_a\times\psi_a}.
\end{equation}
The cost matrix $\bm{C}$ of the joint labeled GCI fusion is given by
\begin{equation}
    \bm{C}=\begin{bmatrix}
    \bm{C}_1 &\bm{C}_2
    \end{bmatrix}
\end{equation}

The $k$ best assignments can be obtained by the Murty's algorithm \cite{Murty1968,Miller1997},
\begin{equation}
\label{eq:AssignmentMurty1}
\begin{gathered}
    \begin{aligned}
    &Assign_1\\
    &=\{(\ell_{\tau_1(1),a},\ell_{\tau_1(1),b}),\ldots,(\ell_{\tau_1(n_1),a},\ell_{\tau_1(n_1),b});cost_1\}\\
    &=\{L_1;cost_1\},
    \end{aligned}\\
    \vdots\\
    \begin{aligned}
    &Assign_k\\
    &=\{(\ell_{\tau_k(1),a},\ell_{\tau_k(1),b}),\ldots,(\ell_{\tau_k(n_k),a},\ell_{\tau_k(n_k),b});cost_k\}\\
    &=\{L_k;cost_k\}
    \end{aligned},
\end{gathered}
\end{equation}
where
\begin{equation}
\begin{aligned}
    &cost_i\\
    &=\sum_{j=1}^{n_i}-\log[ \frac{\eta(\ell_{\tau_i(j),a},\ell_{\tau_i(j),b})r_a(\ell_{\tau_i(j),a})^{\omega_a} r_b(\ell_{\tau_i(j),b})^{\omega_b} }{(1-r_a(\ell_{\tau_i(j),a}))^{\omega_a} (1-r_b(\ell_{\tau_i(j),b}))^{\omega_b} } ].
\end{aligned}
\end{equation}
Let
\begin{equation}
    cost_0=\sum_{i=1}^{\psi_a}\log[ (1-r_a(\ell_{i,a}))^{\omega_a} ] + \sum_{j=1}^{\psi_b}\log[(1-r_b(\ell_{j,b}))^{\omega_b}],
\end{equation}
For any label set $L_i$ in (\ref{eq:AssignmentMurty1}), the corresponding weight $w(L_i)$ is given by
\begin{equation}
    w(L_i)=\exp{\{ cost_0-cost_i \}},
\end{equation}
the normalization coefficient $C$ is given by
\begin{equation}
    C = \sum_{i=1}^kw(L_i),
\end{equation}
the single-target state density $\bar{p}(\cdot)$ is given by (\ref{eq:singleTargetFusion}).

\begin{remark}
If the single-target state density of the LMB RFS is represented by the Gaussian mixture model,
\begin{equation}
    p(x)=\sum_{i=1}^M\alpha_i\mathcal{N}(x;\hat{x}_i,\bm{P}_i),
\end{equation}
then $p(x)^{\omega}$ can be approximated by \cite{Battistelli2013}
\begin{equation}
\begin{aligned}
    &[\sum_{i=1}^M\alpha_i\mathcal{N}(x;\hat{x}_i,\bm{P}_i)]^\omega\approx \sum_{i=1}^M[\alpha_i\mathcal{N}(x;\hat{x}_i,\bm{P}_i)]^\omega_i\\
    &=\sum_{i=1}^M\alpha_i^\omega\kappa(\omega,\bm{P}_i)\mathcal{N}(x;\hat{x}_i,\frac{\bm{P}_i}{\omega}),
\end{aligned}
\end{equation}
where
\begin{equation*}
    \kappa(\omega,\bm{P}_i)\triangleq\frac{[\det{(2\pi \bm{P}_i\omega^{-1})}]^{\frac{1}{2}}}{[\det{(2\pi\bm{P}_i)}]^{\frac{\omega}{2}}}.
\end{equation*}
If the fusion method for two Gaussian mixture probability density is given by \cite{Battistelli2013}

\begin{equation}
\label{eq:solveBarP1}
    \bar{p}(x)=\frac{[p_1(x)]^{\omega}[p_2(x)]^{1-\omega}}{\int[p_1(x)]^{\omega}[p_2(x)]^{1-\omega}dx},
\end{equation}
then
\begin{equation}
\label{eq:solveBarP2}
    \bar{p}(x)=\frac{ \sum_{i=1}^{M_1}\sum_{j=1}^{M_2}\alpha_{ij}\mathcal{N}(x;\hat{x}_{ij},\bm{P}_{ij}) }{ \sum_{i=1}^{M_1}\sum_{j=1}^{M_2}\alpha_{ij} },
\end{equation}
where
\begin{align*}
    &\bm{P}_{ij}=[\omega(\bm{P}_i^1)^{-1}+(1-\omega)(\bm{P}_j^2)^{-1}],\\
    &\hat{x}_{ij}=\bm{P}_{ij}[ \omega(\bm{P}_i^1)^{-1}\hat{x}_i^1+(1-\omega)(\bm{P}_j^2)^{-1}\hat{x}_j^2 ],\\
    &\begin{aligned}
    \alpha_{ij}&=(\alpha_i^1)^\omega(\alpha_j^2)^{1-\omega}\kappa(\omega,\bm{P}_i^1)\kappa(1-\omega,\bm{P}_{j}^2)\\ &\cdot \mathcal{N}(\hat{x}_i^1-\hat{x}_j^2;0,\frac{\bm{P}_i^1}{\omega}+\frac{\bm{P}_j^2}{1-\omega}).
    \end{aligned}
\end{align*}
$\eta(\cdot)$ and $\bar{p}(\cdot)$ can be obtained by applying this method.
\end{remark}

The fused multi-target density is given by
\begin{equation}
    \bm{f}_V(\bm{X})=\Delta(\bm{X})w(\mathcal{L}(\bm{X}))[\bar{p}]^{\bm{X}},
\end{equation}
where $\mathcal{L}(\bm{X})\in\{L_1,\ldots,L_k\}$. Finally, we obtain the LMB RFS $\{(\bar{r}(\ell_a,\ell_b),\bar{p}(x,\ell_a,\ell_b))\}_{\ell_a\in\Psi_a,\ell_b\in\Psi_b}$ by applying
\begin{equation}
    \bar{r}(\ell_a,\ell_b)=\sum_{L\in\{L_1,\ldots,L_k\}}w(L)1_L(\ell_a,\ell_b).
\end{equation}
This approximation can guarantee the invariance of the first order moment of the probability density.

The simplified joint labeled GCI fusion method does not need the cost matrix, given the parameter set $\bm{f}_a(\bm{X})=\{ (r_a(\ell),p_a(x,\ell)) \}_{\ell\in\Psi_a}$ and $\bm{f}_b(\bm{X})=\{ (r_b(\ell),p_b(x,\ell)) \}_{\ell\in\Psi_b}$, the fused multi-target density is given by
\begin{equation}
    \bm{f}_V(\bm{X})=\{ (\bar{r}(\ell_a,\ell_b),\bar{p}(x,\ell_a,\ell_b)) \}_{\ell_a\in\Psi_a,\ell_b\in\Psi_b},
\end{equation}
where
\begin{equation}
\begin{small}
\begin{aligned}
    &\bar{r}(\ell_a,\ell_b)\\
    &=\frac{\eta(\ell_a,\ell_b) r_a(\ell_a)^{\omega_a}r_b(\ell_b)^{\omega_b} }{(1-r_a(\ell_a))^{\omega_a}(1-r_b(\ell_b))^{\omega_b}+\eta(\ell_a,\ell_b)r_a(\ell_a)^{\omega_a}r_b(\ell_b)^{\omega_b}}.
\end{aligned}
\end{small}
\end{equation}

For the convenience of extracting tracks from the LMB RFS, the joint label is marginalized to the standard label,
\begin{equation}
    \bar{r}(\ell_a)=\sum_{\ell_b\in\Psi_b}\bar{r}(\ell_a,\ell_b).
\end{equation}

\subsection{Computational Complexity Analysis}

\begin{table*}[t]
    \centering
    \begin{tabular}{c|c|c|c}
        \hline
         &JL-GCI &Simplified JL-GCI &LM-GCI  \\
         \hline
        Cost Matrix &$\mathcal{O}(L_1L_2M^2)$ &$\mathcal{O}(L_1L_2M^2)$ &$\mathcal{O}(L_{max}^2M^2)$\\
        Assignment &$\mathcal{O}(kL_1^3(L_1+L_2))$ &- &$\mathcal{O}\{ L_{max}^3 \}$\\
        Total &$\mathcal{O}(\max\{ L_1L_2M^2, kL_1^3(L_1+L_2)\})$ &$\mathcal{O}(L_1L_2M^2)$ &$\mathcal{O}(\max\{ L_{max}^2M^2,L_{max}^3 \})$\\
        \hline
    \end{tabular}
    \caption{The computational complexity comparison of the three methods}
    \label{tab:my_label}
\end{table*}

For the fusion of two multi-target probability density functions, suppose that every single-target density is expressed by the Gaussian mixture model of $M$ components, the probability density generated by agent $a$ contains $L_a$ Bernoulli components, and the probability density generated by agent $b$ contains $L_b$ Bernoulli components. For the joint labeled GCI fusion method, the computational complexity for the cost matrix is $\mathcal{O}(L_aL_bM^2)$, that for calculating the optimal $k$ multi-target states is $\mathcal{O}(kL_a^3(L_a+L_b))$. For the simplified method, the computational complexity for computing the LMB RFS is $\mathcal{O}(L_aL_bM^2)$. While the computational complexity of the LM-GCI method is $\mathcal{O}(\max\{ L_{max}^2M^2,L_{max}^3 \})$, where $L_{max}=\max\{ L_a,L_b \}$. In general, the computational complexity of the simplified JL-GCI method is the lowest, that of the JL-GCI method is the highest, while that of the LM-GCI method is between two methods. The computational complexity comparison of the three methods is presented in Table \ref{tab:my_label}. 

\section{Simulation}
\label{Simulation}

The JL-GCI and the simplified JL-GCI are tested in a 2-dimensional multi-target tracking scenario with two sensors, compared with the LM-GCI-LMB method \cite{Li2019}. The single-target density is represented by the Gaussian distribution. The concentration of this paper is not the adaptability of the weights in the GCI fusion, thus the weights are fixed. The LMB RFS for each agent is generated by the LMB filter \cite{Reuter2014}, then delivering to different fusion methods, avoiding the disturbance caused by the multi-target tracking algorithm.

\subsection{TOSPA Metric}
In this paper, we utilize the TOSPA metric \cite{Ristic2010, Mahler2014} to comprehensively evaluate the cardinality and the single-target state estimated by the multi-sensor multi-target fusion method. Given the labeled multi-target state
\begin{equation}
    \begin{aligned}
        \tilde{\bm{X}}&=\{ (x_1,s_1),\ldots,(x_m,s_m) \},\\
        \tilde{\bm{Y}}&=\{ (y_1,t_1),\ldots,(y_n,t_n) \}.
    \end{aligned}
\end{equation}
Without loss of generality, suppose $m\leq n$, the formula of the TOSPA metric is given by
\begin{equation}
\begin{aligned}
    &D_{p,c,\alpha}(\tilde{\bm{X}},\tilde{\bm{Y}})\\
    &=[\frac{1}{n}(\sum_{i=1}^m (d_c(x_i,y_{\pi^*(i)}))^p\\
    &+ \sum_{i=1}^m \alpha^p(1-\delta[s_i,t_{\pi^*(i)}]) + c^p\cdot(n-m) )],
\end{aligned}
\end{equation}
where $\pi^*(\cdot)$ is the optimal assignment for two label sets,
\begin{equation}
    \pi^*=\arg\min_{\pi\in\Pi_n}\sum_{i=1}^m(d_c(x_i,y_{\pi(i)}))^p,
\end{equation}
\begin{equation}
    d_c(x,y)=\min\{c,d(x,y)\},
\end{equation}
\begin{equation}
    d(x,y) = (\sum_{\ell=1}^N|x(\ell)-y(\ell)|^p)^{1/p},
\end{equation}
$N$ is the dimension of single target state, $x(\ell)$ is the $\ell$th element of the vector $x$.

\subsection{Scenario}

The target dynamic is represented by the constant velocity model, the Markov transition density is $f_{k|k-1}(x'|x)=\mathcal{N}(x';\bm{F}x,Q)$, where $\bm{Q}=\bm{G}\bm{G}^T\sigma_{w}^2$, here we choose $\sigma_{w}=10$,
\begin{equation*}
    \bm{F}=\begin{bmatrix}
    1 &T &0 &0\\
    0 &1 &0 &0\\
    0 &0 &1 &T\\
    0 &0 &0 &1
    \end{bmatrix},\quad
    \bm{G} = \begin{bmatrix}
    \frac{T^2}{2} &0\\
    T &0\\
    0 &\frac{T^2}{2}\\
    0 &T
    \end{bmatrix}.
\end{equation*}
The measurement model (likelihood function) is $L_z(x)=\mathcal{N}(z;Hx,R)$, where
\begin{equation*}
    \bm{H}=\begin{bmatrix}
    1 &0 &0 &0\\
    0 &0 &1 &0
    \end{bmatrix},\quad
    \bm{R}=\begin{bmatrix}
    \sigma_x^2 &0\\
    0 &\sigma_y^2
    \end{bmatrix}.
\end{equation*}
In this paper, we choose the standard deviation for the first agent by $\sigma_x=10,\sigma_y=10$, that of the second agent by $\sigma_x=12,\sigma_y=12$. The clutter is Poisson with expectation value $\lambda = 10$ and the spacial distribution is uniform. The probability detection $p_D$ is set to 0.98, 0.88, 0.78 respectively. The target survival probability $p_S$ is set to 0.90. The newborn targets follow the LMB density with the parameter set: $r_1=0.07$,$p_1(x)=\mathcal{N}(x;\hat{x}_1,\bm{P}_1)$;$r_2 = 0.07$,$p_2(x)=\mathcal{N}(x;\hat{x}_2,\bm{P}_2)$;$r_3=0.07$,$p_3(x)=\mathcal{N}(x;\hat{x}_1,\bm{P}_3)$;$r_4 = 0.07$,$p_4(x)=\mathcal{N}(x;\hat{x}_4,\bm{P}_4)$, where $\hat{x}_1=[0,0,0,0]^T$,$\hat{x}_2=[400,0,-600,0]^T$,$\hat{x}_3=[-800,0,-200,0]^T$,$\hat{x}_4=[-200,0,800,0]^T$,$\bm{P}_1=\bm{P}_2=\bm{P}_3=\bm{P}_4=diag([1,1,1,1]\cdot 50^2)$.

\begin{figure}
    \centering
    \includegraphics[width=0.5\textwidth]{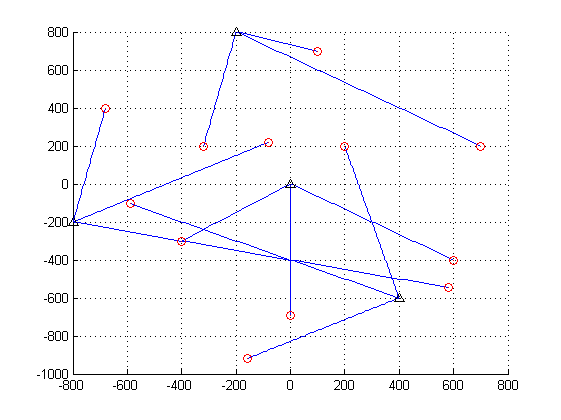}
    \caption{The tracks in the 2-dimensional space. There are 12 targets in the scenario. There is a maximum of 10 targets at a single moment. The scenario includes the birth and disappearance of targets, i.i.d. newborn targets, overlapping of targets. The black triangle indicates where the target appears, and the red circle indicates where the target disappears.}
    \label{fig:pic2}
\end{figure}

The multi-target tracking scenario is referred to \cite{Reuter2014, Vo2013}. The tracks in the 2-dimensional space are presented by Fig. \ref{fig:pic2}. The scenario includes the appearance and disappearance of targets, i.i.d. newborn targets, and the overlapping targets. This scenario challenges the abilities of label management, target tracking, the assignment between multi-sensors, and multi-sensor multi-target fusion.

\subsection{Results}

\begin{figure}
    \centering
    \includegraphics[width=0.5\textwidth]{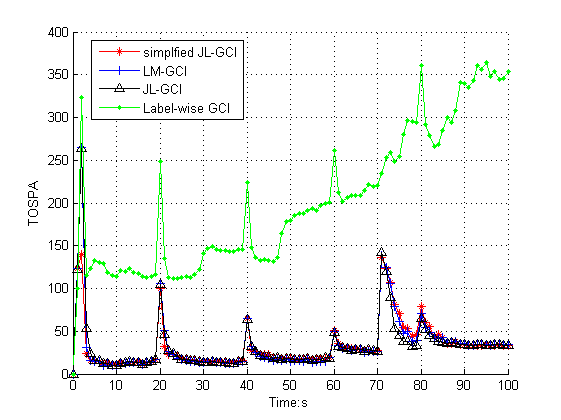}
    \caption{TOSPA for 50-times Monte-Carlo simulation under $p_D=0.98$. The data of the label-wise GCI method presents the label inconsistency for different agents in the scenario, showing excellent performance for the proposed methods and LM-GCI.}
    \label{fig:fig1_1}
\end{figure}

\begin{figure}
    \centering
    \includegraphics[width=0.5\textwidth]{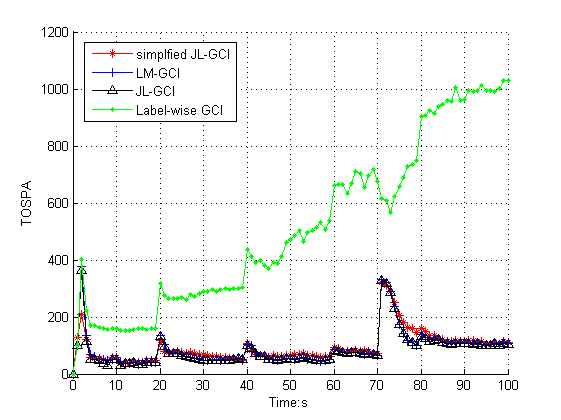}
    \caption{TOSPA for 50-times Monte-Carlo simulation under $p_D=0.88$. Compared with that under $p_D=0.98$, the TOSPA values for the three methods all increase, which may be caused by the drop in performance of each agent's tracker.}
    \label{fig:fig2_1}
\end{figure}

\begin{figure}
    \centering
    \includegraphics[width=0.5\textwidth]{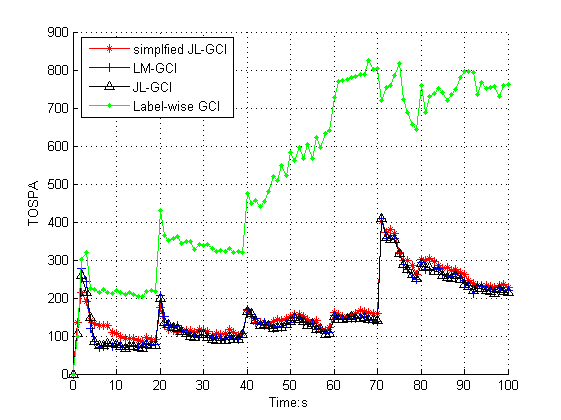}
    \caption{TOSPA for 50-times Monte-Carlo simulation under $p_D=0.78$.}
    \label{fig:fig3_1}
\end{figure}

The performances of the proposed methods under $p_D=0.98,0.88,0.78$ for 50-times Monte-Carlo simulation in terms of TOSPA are presented in Fig. \ref{fig:fig1_1}, \ref{fig:fig2_1}, \ref{fig:fig3_1} respectively. The data of the label-wise GCI method presents the label inconsistency for different agents in the scenario, showing excellent performance for the proposed methods and LM-GCI. Besides, even though the simplified JL-GCI method is supposed to work given the Assumption \ref{Assumption1}, which is not satisfied in this scenario. The simplified method can still achieve an acceptable performance.
\begin{table}[]
    \centering
    \begin{tabular}{c|c|c|c}
        \hline
        $p_D$ &JL-GCI &simplfied JL-GCI &LM-GCI  \\
        \hline
        0.98 &\textbf{32.204} &32.3661 &32.7274 \\
        0.88 &\textbf{84.3894} &92.744 &86.7796 \\
        0.78 &\textbf{160.5579} &173.6075 &162.2721\\
        \hline
    \end{tabular}
    \caption{The average TOSPA for 50-times Monte-Carlo simulation.}
    \label{tab:TOSPA}
\end{table}

\begin{figure}
    \centering
    \includegraphics[width=0.5\textwidth]{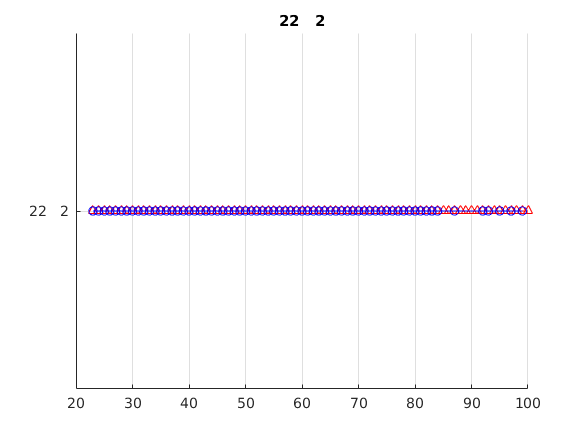}
    \caption{The assignment for the label $(22,2)$ of agent $a$ in a single simulation iteration. The JL-GCI is compared with the LM-GCI, where the blue circle is the assignment for the LM-GCI, while the red triangle is the assignment for the JL-GCI, in which the label for agent $b$ is selected by maximization of weights of joint labels. During the simulation period, the target with a weight larger than 0.5 is chosen, thus causing the target state with a small weight fail to be extracted from the LMB RFS.}
    \label{fig:singleFig1}
\end{figure}

The average TOSPA for three methods is present in Fig. \ref{tab:TOSPA}. The statistical result shows that the JL-GCI method outperforms its simplified version and LM-GCI under different $p_D$. We believe that the better performance of the JL-GCI in this paper is mainly because the method has a more accurate cardinality estimation of the fused multi-target density. In a single simulation iteration, for the same label, the assignment of the JL-GCI and LM-GCI is shown in Fig. \ref{fig:singleFig1}. Here the blue circle is the assignment for the LM-GCI, while the red triangle is the assignment for the JL-GCI. Fig. \ref{fig:singleFig1} shows that the LM-GCI may fail to report the single-target state since different fusion methods generate different weights of the same Bernoulli component. Other situations that may affect the result of state estimation, as the error of the label assignment and the label inconsistency, are not the main reason.
\begin{figure}
    \centering
    \includegraphics[width=0.5\textwidth]{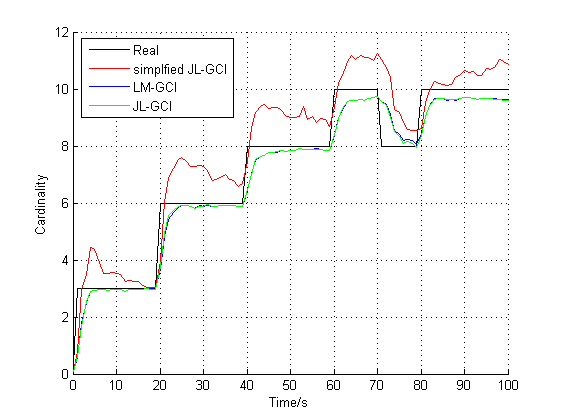}
    \caption{The cardinality estimation for 50-times Monte-Carlo simulation under $p_D=0.98$.}
    \label{fig:figCardBias}
\end{figure}

Fig. \ref{fig:figCardBias} proposes the average cardinality estimation over 50-times Monte-Carlo simulation. Due to introducing the invalid components, the cardinality estimation for the simplified JL-GCI is usually bigger than the real cardinality. For the average cardinality bias over the simulation period, the bias for LM-GCI is $0.3807$, while the bias for JL-GCI is $0.3718$. In a statistical sense, the cardinality estimation of the JL-GCI is more accurate than that of the LM-GCI.

\section{Conclusion}
\label{Conclusion}

Compared with previous methods for label inconsistency, we used the joint label set induced by the direct product for all label spaces, hence the target label for each agent and the label matching among different agents can be represented by the element in the joint label space. On this basis, we obtained the fused joint labeled multi-target probability density by minimizing the weighted KL divergence.

Specifically, we implemented a multi-sensor fusion algorithm for LMB RFSs as well as a simplified version given that the targets are well-separated. Simulations show that the performance of JL-GCI is better than that of the LM-GCI, which is embodied in that the JL-GCI method is more accurate for the cardinality estimation. Compared with the LM-GCI algorithm, the simplified JL-GCI reduces the computational complexity by introducing the well-separated assumption. Simulations show the applicability of the simplified method, even when the assumption does not fully hold.


%

\ifCLASSOPTIONcaptionsoff
  \newpage
\fi

\end{document}